# A Disordered Photonic Medium Enabling Ultrabroadband Light Scattering and Selective Longwave Infrared Emission

Zhenpeng Li[1], Mathis Degeorges[1], Nithin Jo Varghese[1], Jyotirmoy Mandal[1,2]*

[1]Department of Civil and Environmental Engineering, Princeton University, Princeton, USA

[2]Princeton Materials Institute, Princeton University, Princeton, USA

*Corresponding author: Jyotirmoy Mandal (jm3136@princeton.edu)

## Abstract

A surface that selectively emits heat in the long-wave infrared (LWIR) can enable passive cooling in hot environments while retaining partial radiative insulation in cold conditions. However, its cost-effectiveness, practical deployment, and fundamentally, the optical design, remain limited by the reliance on metal reflectors. To overcome this limitation, here we use an absorption-scattering competition factor to establish design guidelines for enhancing reflection or absorption in disordered media across the ultrabroadband ultraviolet-to-far-infrared range. Based on electromagnetic simulations and the optical constants of real materials, we then propose disordered photonic media with a layered, multiscale scattering architecture which, unlike typical scattering designs, simultaneously attains ultraviolet-to-far-infrared reflection and selective LWIR emission. We validate this approach by developing a metal-free selective emitter that exhibits high LWIR emittance (0.88), strong solar reflectance (0.97), and low thermal emittance outside the LWIR (0.49), independent of substrates. Field tests, supported by theoretical modelling, show both enhanced radiative cooling and seasonal thermoregulation performance relative to a state-of-the-art broadband radiative cooler. By expanding the spectral functionality of disordered scattering media as a scalable and low-cost optical materials platform across the solar-to-thermal infrared waveband, this work provides a pathway towards improved energy savings and thermal comfort through passive radiative thermal management.

## Introduction

In recent decades, the field of thermal photonics has seen significant advancements on spectral and angular control of radiative heat flows for various thermal and energy applications.[1–3] A notable example is passive daytime radiative cooling, which ideally requires materials with high reflectance at $\lambda$~0.3–8 μm and ~13–30 μm, and a selective wideband emittance in the longwave infrared (LWIR, $\lambda$~8–13 μm) waveband (Fig. S1).[4,5] In hot environments, a surface with such optical properties is radiatively shielded from the sun and quasi-isolated from both atmospheric and terrestrial radiation, but can still access the heat sink of outer space through the atmosphere's LWIR transmission window, spontaneously cooling both the surface and the Earth.

However, achieving such spectral selectivity over the solar and thermal infrared (TIR) waveband, which spans two orders of magnitude in wavelength ($\lambda$~0.3–30 μm), is challenging. Of the selective LWIR emitters that have been investigated since the 1960s,[6,7] the overwhelming majority employ an optical architecture where a LWIR-emissive material with controlled thickness is placed over an metallic reflector, such as aluminum and silver (Tables S1-S2).[4,7–13] The handful of exceptions employ bulk ceramics with suitable Reststrahlen reflectance bands to achieve selective emittance.[12,14,15] These architectures have been long-known, but face limitations. For instance, the corrodibility, specular glare, impermeability, and intrinsic limits on the solar reflectance of metal films could hinder their practical performance and widespread deployment.[16–18] The few suitable Reststrahlen-reflective ceramics are less practical still, as they are costly to fabricate at scale and are susceptible to chemical weathering.[19,20]

Despite the excellent optical performance of various engineered thermal photonic structures, scalability and material cost remains primary bottlenecks to their real-world impact.[1] To overcome this, radiative cooler (RC) design has exploited disordered photonics featuring random structures to enable multiple scattering of sunlight.[21,22] A variety of disordered media, including porous polymers,[23–25] ceramics,[26,27] and paints,[28,29] now achieve solar reflectances that surpass those of metal mirrors. The inherent scalability and low cost have positioned them as the dominant technology in commercial RC products. Yet, solar-transparent dielectrics used in these RCs typically have broad TIR absorption, yielding broadband TIR emittance. Selectively LWIR-absorbing variants[13,30,31] also function as broadband RCs when applied onto non-metal substrates, with TIR emittances of ~0.84-0.97.[32] While effective at cooling sky-facing surfaces at near or above ambient temperatures,[5,33] broadband emittance may limit subambient cooling applications. More importantly, for vertical surfaces like facades, broadband RCs undergo excessive radiative heat exchange with the terrestrial environments, which can be a major detriment in both hot urban settings and cold climates.[12,34,35]

Beyond making selective LWIR emitters more practical, determining to what extent ultrabroadband reflection is realistically possible by disordered photonic structures, and whether that could be coupled with intrinsic absorption of constituents to achieve selective emittance without the use of bulk metal or Reststrahlen reflectors, remains an important question for advancing radiative thermoregulation and the optical design of scattering media.

In this work, we investigated the design principles for disordered scattering media across the ultraviolet-to-far-infrared (UV-FIR) waveband, targeting optical outcomes of maximizing either diffuse reflectance or absorptance. Our study shows that typical design paradigms, and most established selectively LWIR emissive materials, cannot yield a scattering media acting as substrate-independent selective LWIR emitter. For real materials, that requires specific sizes and spatial distributions of scatterers, and intrinsic properties of material constituents. Guided by our findings, we propose disordered media featuring layered, multiscale light scatterers as a promising optical platform to simultaneously achieve ultrabroadband solar-to-TIR reflectance and selective emittance in the intermediate band (**Figure 1**a). Based on a theoretically guided optical design, we develop a selective LWIR emitter consisting of a nanoporous poly(4-methyl-1-pentene) (nPMP) bonded to a microfibrous polypropylene (μPP) underlayer, which achieves a notable solar reflectance ($R_{solar}$=0.97±0.01), LWIR emittance ($\varepsilon_{LWIR}$=0.88±0.02), and LWIR selectivity ($\eta=\varepsilon_{LWIR}/\varepsilon_{TIR}$=1.42±0.04) (**Figure 1**b). To our knowledge, this nPMP-μPP design is the first scattering and first polymeric media to achieve such selectivity without the aid of broadband metallic or ceramic reflectors (**Figure 1**c, Tables S1 and S2).

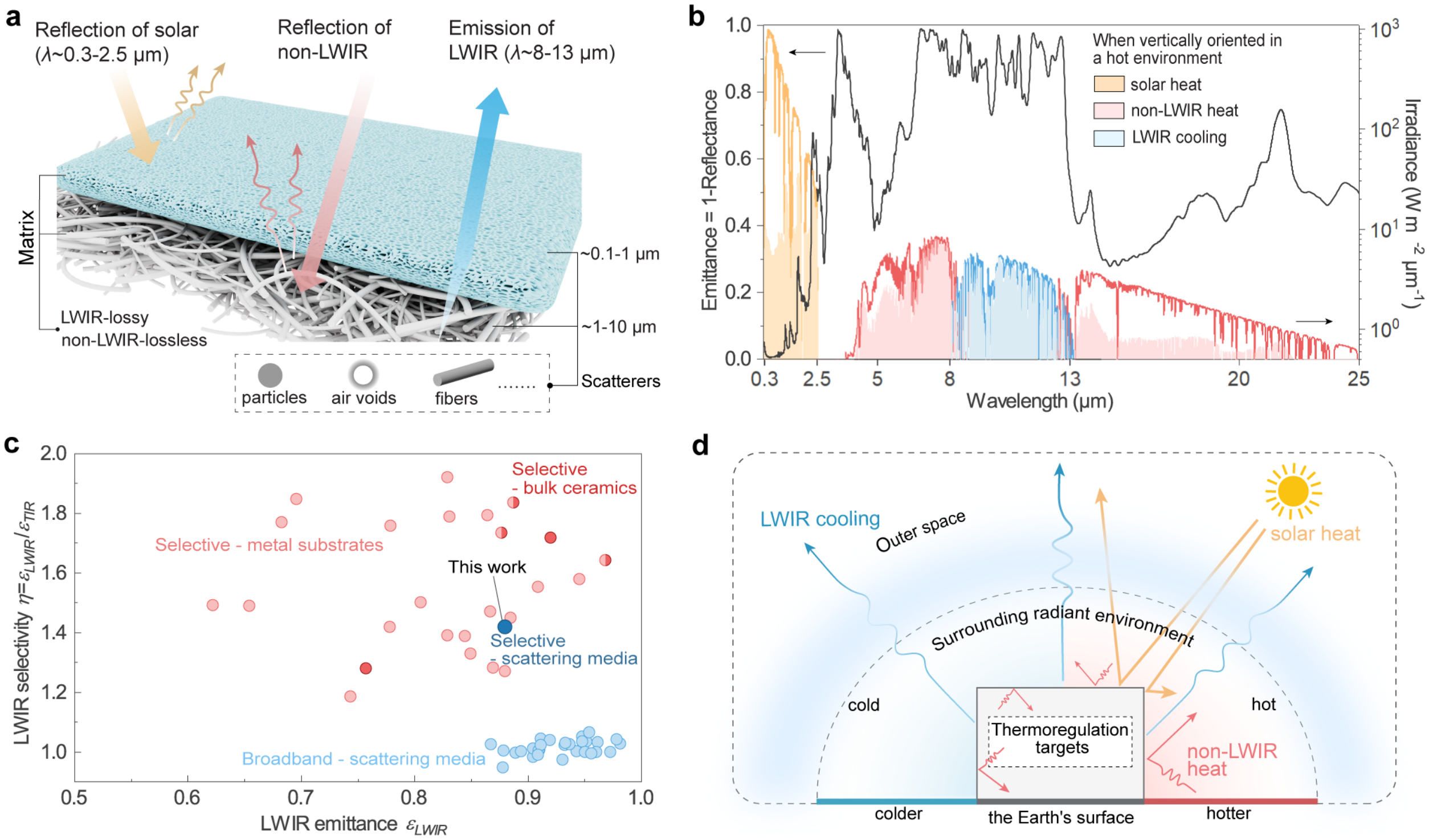

**Figure 1. An ultrabroadband scattering medium with selective LWIR emission for radiative thermoregulation.** (**a**) Schematic diagram of the proposed scattering medium, featuring a layered, multiscale structure to achieve ultrabroadband ultraviolet-to-far-infrared reflectance and selective long-wave infrared (LWIR) emittance. (**b**) Spectral emittance (=1-reflectance) of the nPMP-μPP selective emitter developed in this work. The shaded regions indicate how the nPMP-μPP, when vertically oriented, interacts with a hot radiant environment in the solar and TIR wavelengths compared to an ideal black broadband absorber/emitter (represented by solid lines): rendering solar irradiance intensity to the order of thermal irradiance; reducing non-LWIR heat gains; allowing LWIR cooling. (**c**) Comparisons of LWIR emittance $\varepsilon_{LWIR}$ and LWIR selectivity $\eta=\varepsilon_{LWIR}/\varepsilon_{TIR}$ between the nPMP-μPP and other radiative cooling materials reported in literature (Tables S1 and S2). (**d**) Compared to broadband emitters, selective LWIR emitters maintain radiative cooling to the sky, while blocking unwanted non-LWIR radiative heat gains or losses due to temperature variations of the surrounding environments. This can be used to thermoregulate targets such as buildings and human beings.

Through field tests validated against heat transfer modelling, we further demonstrate our design's passive radiative thermoregulation abilities relative to conventional solar-reflective, broadband-emissive radiative coolers for sky-facing and vertical orientations, and across scenarios. This arises from its low non-LWIR emittance which reduces unwanted radiative heat gains or losses driven by temperature variations of its outdoor radiant environment (**Figure 1**b and d). Deployed as a sky-facing surface, nPMP-μPP reaches lower sub-ambient temperatures than a broadband RC with near-identical $R_{solar}$ and $\varepsilon_{LWIR}$, under intense sunlight, limited atmospheric transparency, and strong convective heat transfer. For vertical surfaces during both hot summer daytime and cold nighttime conditions, the nPMP-μPP reduces non-LWIR radiative heat gains or losses by ~10–30 W m$^{-2}$, demonstrating a passive, seasonal thermoregulation ability. Thermal and energy simulations indicate that, compared with broadband RCs, walls clad with nPMP-μPP may enable cross-seasonal building energy savings comparable to those achieved by painting dark roofs white. If used as architectural fabrics, nPMP-μPP may reduce heat gains into enclosed sheltered spaces by ~45%–90% relative to conventional white fabrics in hot daytime and heat losses by ~25% in cold nighttime.

Promisingly, established fabrication and tunability of the scattering-based design make it compatible with industry-scale production at low material cost (~US$3 m$^{-2}$), and provides a platform for tailoring solar reflectance and thermal emittance to applications in different climates. Together, these results show a practical way to harness optical scattering media for improved passive thermoregulation of buildings, shelters, and human beings. More fundamentally, our study presents a new paradigm for disordered scattering media design targeting ultrabroadband spectral control.

## Optimizing Light Scattering for Selective LWIR Emittance

The fundamental idea that disordered media can enhance reflectance through multiple scattering has been long known.[36] White paints, for instance, employ random titania particles with ~0.2-1.0 μm sizes to achieve high visible reflectance. More recently, RCs have used nanoscale random scatterers, or hierarchies of nano- and microscale scatterers to strongly reflect sunlight.[13,23,26–28] Since scattering intensity depends on the size parameter $d/\lambda$, where $d$ is the particle size and $\lambda$ is light wavelength, it follows that upscaling the scatterers with suitable materials should extend the reflectance from solar to TIR. However, there are fundamental limitations because the form factors of realistic scatterers—specifically their size parameters and optical extinction coefficients $\kappa$—exhibit strong spectral dependence across disparate wavelength scales, making the design of scattering media for realizing ultrabroadband reflectors and selective emitters a non-trivial task.

### *Scattering-Absorption Competition in Scattering Media*

To optimize reflectance $R(\lambda)$ or absorptance $A(\lambda)$ ($A$=$\varepsilon$) over an ultrabroad waveband, we used the absorption-scattering competition factor $\zeta$ to aid designing scattering media, which quantifies the tendency of incident light to be absorbed before being diffusely transported via multiple scattering. For an individual scatterer, $\zeta = Q_{abs}/[Q_{sca}(1-g)]$, where $Q_{abs}$ is absorption efficiency, $Q_{sca}$ is scattering efficiency, and $g$ is asymmetry parameter (Note S1). **Figure 2**a shows $\zeta$ as the function of $d/\lambda$ for spherical scatterers with refractive indices $\tilde{n} = 1.5 + \kappa i$. When $\kappa = 0$, $\zeta = 0$ regardless of $d/\lambda$. This implies that a sufficiently thick ensemble of lossless scatterers would attain ideal $R(\lambda)$ when $\lambda \lesssim d$, as they strongly scatter wavelengths comparable to (via Mie scattering) or significantly smaller than their sizes (via geometric scattering).

In reality, materials have non-zero optical loss due to electronic transitions and vibrational resonances, especially in the UV and TIR.[37,38] For lossy materials, $\zeta$ rapidly increases from its minimum in the Mie regime towards both geometric and Rayleigh regimes, roughly by an order of magnitude for a similar change in $d/\lambda$ (**Figure 2**a). In the geometric regime, this arises from scattering being mostly forward and $Q_{abs}$ increases with $d$, while in the Rayleigh regime this arises from $Q_{sca}$ being much weaker. For refractive indices with higher real parts, e.g., $\tilde{n} = 2 + \kappa i$, very similar trends are observed (Fig. S2). The results have two implications – firstly, for a given lossy particle, the most efficient scattering relative to loss occurs in a limited range where $0.3d \lesssim \lambda \lesssim d$ i.e., it is not ultrabroadband. Secondly, the rapid rise in $\zeta$ away from the Mie regime indicates that for $\lambda \gg d$ or $\lambda \ll d$, absorption becomes more significant in scattering media.

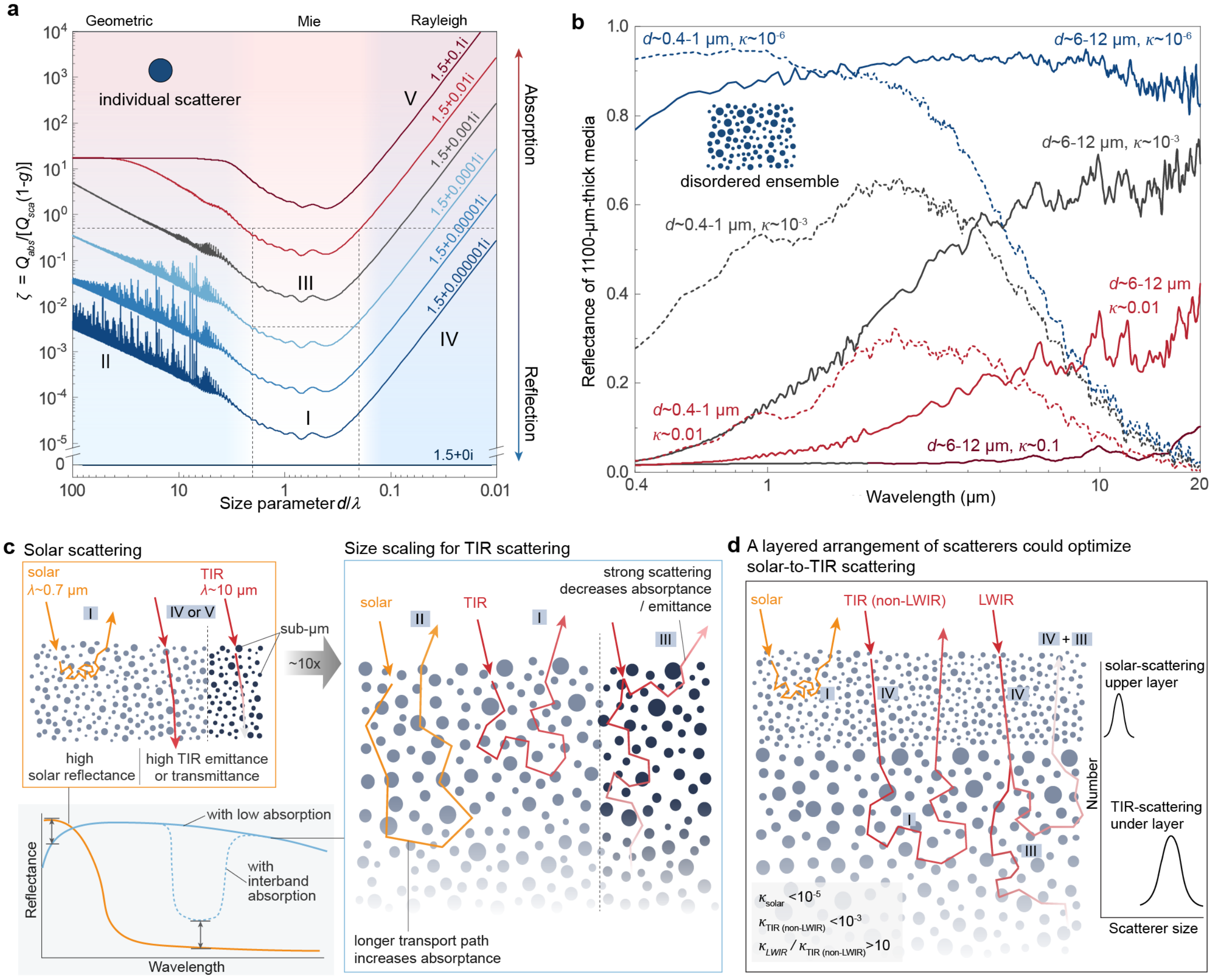

**Figure 2. Optimizing Light Scattering for Selective LWIR Emittance.** (**a**) Calculated ratio $\zeta$ of absorption efficiency to reduced scattering efficiency as a function of the size parameter $d/\lambda$ for scatterers with $n$=1.5 and different extinction coefficients $\kappa$, spanning the geometric, Mie and Rayleigh regimes. Distinct optical regimes (I–V) are highlighted. (**b**) Simulated spectral reflectance of 1100-$\mu$m-thick disordered scattering media with different sizes of scatterers and for different $\kappa$, with air being the low-index component. (**c**) Schematic illustration of photon transport in solar-scattering media and TIR-scattering media. Conventional solar-scattering media based on submicron features exhibit high solar reflectance but large TIR transmittance or emittance. Increasing the characteristic scatterer size by approximately one order of magnitude shifts the scattering window towards the TIR but could not achieve optimal solar reflectance and LWIR emittance. (**d**) Schematic of a layered disordered-scattering architecture for optimized solar-to-TIR scattering. A solar-scattering upper layer composed of smaller scatterers is combined with a TIR-scattering underlayer composed of larger scatterers, enabling strong solar reflection, and selective LWIR emission when the material exhibits low $\kappa$ outside the LWIR but higher $\kappa$ within the LWIR.

The significance of $\zeta$ when $\kappa \neq 0$ for scattering media becomes evident from finite-difference time-domain (FDTD) simulations of separate ensembles of large particles ($d{\sim}6-12$ μm) and of small particles ($d{\sim}0.4-1$ μm), using different $\tilde{n} = 1.5 + \kappa i$ values across $\lambda = 0.4-20$ μm (**Figure 2**b). For these non-dilute scattering media with a scatterer volume fraction of ~0.3, $R(\lambda)$ is highest in the Mie regime, rapidly decreases as expected towards the Rayleigh regime, and notably, also drops substantially towards the geometric regime even when $\kappa$ is as low as $10^{-6}$. Both the maximum value and the bandwidth over which $R(\lambda)$ is near its maximum decrease significantly with $\kappa$, in agreement with the behavior predicted by $\zeta$ for individual scatterers. Collectively, the results in **Figure 2**a-b establish a clear connection between the $\zeta$ of individual scatterers and the $R(\lambda)$ of their disordered ensembles. The peak $R(\lambda)$ of scattering media with different $\kappa$ suggest that $\zeta{\sim}10^{-4}$, $10^{-2}$ and 1 correspond to good, modest and low reflectance, respectively. Conversely, the increasing minimum of $\zeta$ with $\kappa$ provides an estimate of the $\kappa$ required to obtain high (~0.9), moderate (~0.5) and low ($\lesssim$ 0.1) reflectance through multiple Mie scattering, i.e., $\sim10^{-6}$, $\sim10^{-3}$, $\sim10^{-1}$, respectively.

### *Intuitive Size-Scaling Compromises Ultrabroadband Performance of Real Materials*

For solar-scattering designs, $\zeta$ has typically not been a critical consideration. This is partly because ~80% of solar heat lies within a relatively narrow waveband ($\lambda{\sim}0.4-1.4$ μm). When using polydisperse, sub-μm scatterers,[22,28,30] their sizes are comparable to $\lambda$ such that scattering occurs mainly in the Mie regime where $\zeta$ approaches the lowest. In addition, many dielectric materials exhibit very low $\kappa$ ($\lesssim 10^{-6}$) across this waveband,[23,37,39,40] which keeps $\zeta$ very low. Consequently, even if $\zeta$ is sub-optimal near the solar waveband edges, high $R_{solar}$ has been widely achieved in scattering media. In the TIR, however, where $\lambda \gg d$, such effective solar scatterers become nearly non-scattering and thus the media naturally transition into broadband emitters or transmitters, depending on the value of $\kappa$ in the TIR (**Figure 2**c).[23,26–28,30,41–43]

To extend the high $R(\lambda)$ from the solar to the TIR band as a precondition for selective LWIR emittance, it is necessary to increase $d$ by severalfold to an order of magnitude while keeping $\kappa$ sufficiently low at non-LWIR wavelengths, such that $\zeta$ remains low there (**Figure 2**c). However, the extreme breadth of the solar-to-TIR waveband, and the rarity of materials with suitable $\kappa$ across it, means that intuitive scaling without careful consideration of $\zeta$ can greatly compromise performance in critical parts of the UV-to-FIR waveband.

One major limitation arises in solar band, particularly in the UV-visible range. As evident from **Figure 2**a, $\zeta$ increases by >10 times from its minimum when $\lambda \lesssim 0.1d$. In such condition, weaker and more forward-directed scattering prolongs the transport path of shortwave light within the medium (**Figure 2**c), enhances $A(\lambda)$, and thus appreciably reduces $R_{solar}$ even when the medium is sufficiently thick (**Figure 2**b and Note S1). This behavior is further corroborated by comparing the reflection spectra of nano- and micro-structured forms of materials known to have very low $\kappa$ across $\lambda{\sim}0.3-1$ μm (e.g., $Al_2O_3$, MgO, and polypropene, Fig. S3). Thus, sub-μm scatterers do not effectively reflect TIR radiation ($\lambda{\sim}10$ μm), whereas micro-sized scatterers are likewise suboptimal for reflecting short-wavelength sunlight ($\lambda{\sim}0.3-0.7$ μm). Although this optical mechanism has seldom been articulated, similar spectral behavior is also evident in prior reports.[44–47]

Another critical limitation arises in achieving spectrally selective emittance, here within the LWIR band. Large scatterers used to efficiently scatter and reflect TIR radiation on either side of the LWIR band also Mie-scatter LWIR light, as $d{\sim}\lambda_{LWIR}$. Thus, $Q_{sca}$ remains sufficiently high that $\zeta$ cannot approach the absorption-dominant regime ( $\zeta > 1$ ) unless $\kappa \gtrsim 0.1$ (**Figure 2**a). However, this requirement, together with the need to achieve better-than-moderate $R(\lambda_{non\text{-}LWIR})$ (i.e., $\zeta < 10^{-2}$), implies that $\kappa$ must increase sharply by $\sim 10^3$ at the non-LWIR to LWIR transition points. *To our knowledge, no such materials exist.* We are therefore limited to two sub-optimal cases. If non-LWIR $\kappa$ is kept low (< $10^{-3}$) to enable strong reflectance, then $\kappa$ in the LWIR is also too low (< 0.1) to suppress scattering, and $\varepsilon_{LWIR}$ is limited by near-surface backscattering (**Figure 2**c). Conversely, if scatterers with $\kappa_{LWIR} \gtrsim 0.1$ are used to yield high $\varepsilon_{LWIR}$, then $\kappa$ near the LWIR edges is also too large, and $R(\lambda_{non\text{-}LWIR})$ is comprised. In short, high spectral selectivity is unlikely to be achieved using size-scaled disordered scattering media alone without careful design of both the materials and the scattering architecture.

### *Mixing Scatterers of Different Size Scales Also Yields Suboptimal Performance*

As indicated in **Figure 2**, size distributions in the sub-μm or $\sim$10 μm range are individually insufficient given how their $\zeta$ varies under non-zero optical loss. Yet, such small and large scatterers remain necessary for efficient solar and TIR reflection, respectively. We thus considered whether a homogeneous mixture of scatterers with such disparate size distributions can overcome the limitations associated with a single-scale size range. However, FDTD simulations of scatterers made of hypothetical selective LWIR-emissive materials indicate that such mixing remains suboptimal (Fig. S4). For example, such mixing markedly reduces $R(\lambda_{non\text{-}LWIR})$ compared with the micro-scale scattering medium. This is because the sub-μm scatterers filling the space between the large ones contribute little to TIR scattering, but instead form an effective lossy medium with lower index contrast with the large scatterers than air. This suppresses TIR scattering and enhances absorption, increasing not only $\varepsilon_{LWIR}$ but also $\varepsilon$ across the TIR, thus rendering selective LWIR emittance. As shown later, mixing also yields a sub-optimal $R_{solar}$, between that of sub-μm scattering medium and the less optimal micro-scale scattering medium.

Notably, this result stands in contrast to the understanding that mixing scatterers with a wide range of sizes around $d \sim \lambda_{target}$ may yield high reflectance at $\lambda \sim \lambda_{target}$. Such an approach has long been utilized to realize high $R_{solar}$ in scattering media with ultralow $\kappa$.[48,49] Here we show that, when $\kappa > 0$, extending this strategy to ultrabroad solar-to-TIR bands through random mixtures of comparably broad size distributions may be suboptimal.

### *Proposed Design: Layering Scatterers of Different Size Scales*

These considerations led us to consider a disordered photonic medium, in which scatterers with different size ranges are spatially separated along the depth. When targeting the ultrabroad solar-to-TIR wavelengths that span about two orders of magnitude, at least two layers are needed: an upper layer comprising sub-μm scatterers and an underlayer comprising micro-sized scatterers (**Figure 2**d). For selective LWIR emitters targeted here, the materials for the scatterers must have very low $\kappa$ (<$10^{-3}$) except in the LWIR. With this optical architecture, the underlayer is expected to provide broadband reflection across the solar-to-TIR wavelengths, which is essential for achieving selective LWIR emission without relying on any reflective substrate, and also functions as a moderate solar reflector and LWIR emitter. The sub-μm scatterers in the top layer would more efficiently reflect short-wavelength light to enhance $R_{solar}$ and, because this layer is a very weakly TIR-scattering yet intrinsically LWIR-absorptive effective medium, can also significantly add to the $\varepsilon_{LWIR}$ of the underlayer without comprising non-LWIR reflectance.

## Material Selection and Numerical Validation of the Design Approach

Having hypothesized the general design paradigm, we then validated it through simulations and experiments based on real materials, the first of which is detailed below.

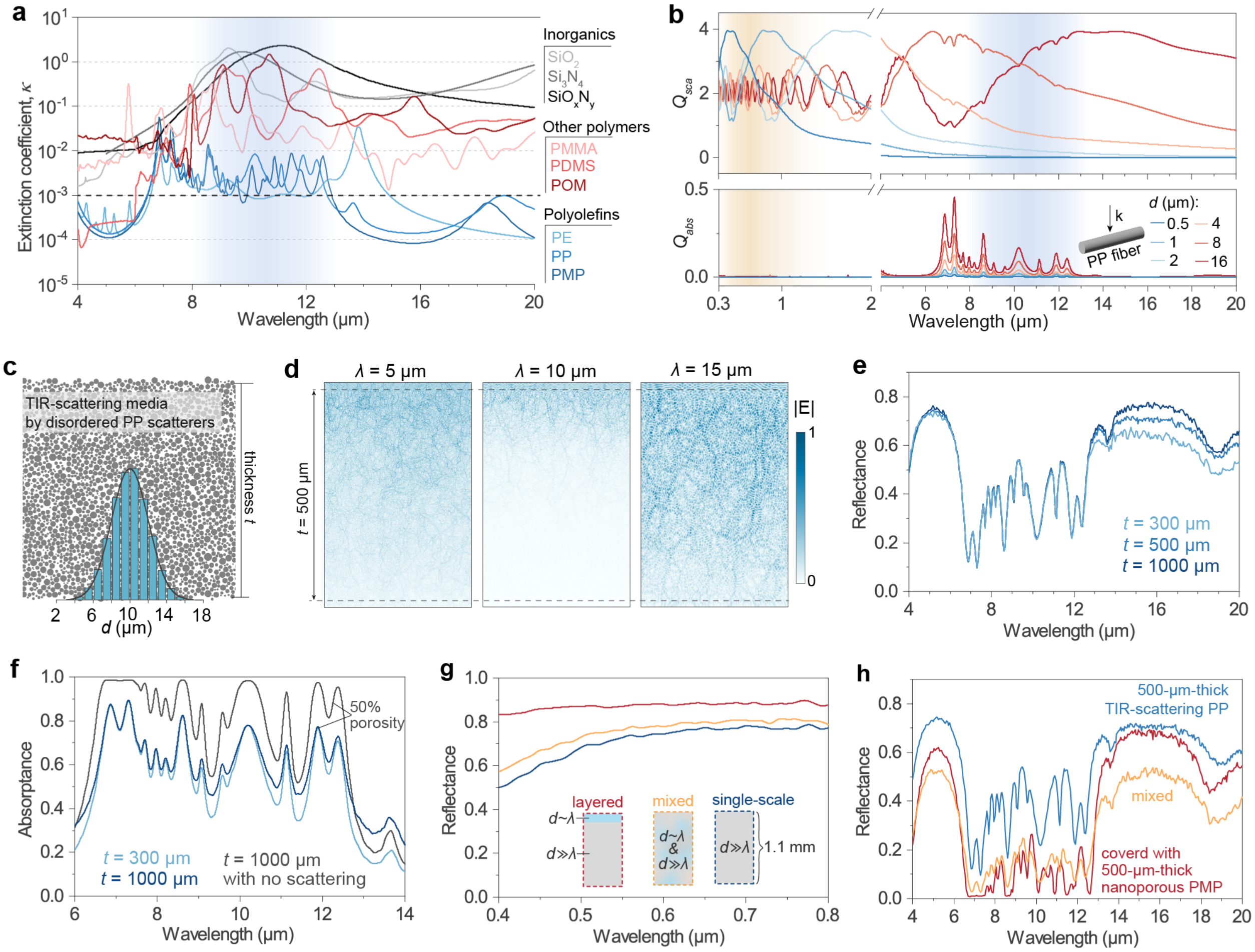

**Figure 3. Material selection and numerical validation.** (**a**) The spectral extinction coefficients of some common inorganics and polymers that have been utilized to realize selective LWIR emitters, in comparison with polyolefins including polyethylene (PE), polypropylene (PP), and poly(4-methyl-1-pentene) (PMP). (**b**) Scattering efficiency $Q_{sca}$ and absorption efficiency $Q_{abs}$ of PP fibers with different diameters $d$. (**c**) Visualization of a random packing of polydisperse scatterers, which is used for simulating the disordered scattering medium having scatterers with an average diameter of 10 μm and a volume fraction of 0.5. Insert: particle size distribution diagram. (**d**) Simulated distribution of normalized electric fields within the PP-based TIR-scattering medium under plane-wave illumination at wavelengths of 5 μm, 10 μm and 15 μm. (**e**) Simulated reflection spectra of the PP-based TIR scattering medium at thicknesses of 300, 500, and 1000 μm. (**f**) Simulated absorption spectra of the TIR-scattering medium made with PP, in comparison with a 1000-μm-thick effective medium with the same volume fraction of PP but exhibits no light scattering. (**g**) Simulated reflection spectra in visible waveband for a thick scattering medium with different configurations of scatterers (Fig. S13). (**h**) Simulated reflection spectra in TIR waveband for a 500-μm-thick PP TIR-scattering medium covered with/without a 500-μm-thick nanoporous PMP upper layer, as well as a 1000-μm-thick medium where sub-μm PMP and micro-sized PP scatterers are mixed (Fig. S14).

### *Material Selection: Criteria for κ Rules Out Most Known Selective Emitters*

Although the size and position of scatterers could be controlled through fabrication, satisfying the requirement for $\kappa$ first requires the selection of appropriate materials. Non-metallic TIR reflectors are, in fact, much rarer than solar reflectors, largely because most materials transparent to solar radiation are intrinsically absorptive in the TIR.[37] A few exceptions, notably, dielectrics such as $Si_3N_4$, $SiO_xN_y$, polyoxymethylene (POM), and poly(methyl methacrylate (PMMA) have been used to produce some of the most selective LWIR emitters reported to date, yet their spectral $\kappa$, while high ($\gtrsim 1$) in the LWIR, also exhibits substantial tails outside ($\kappa \gtrsim 0.1$) (**Figure 3**a).[38,50,51] Consequently, these materials can be selectively LWIR emissive as thin films on metals, but once structured into TIR-scattering media, their high $\kappa$ and $\zeta$ make absorption dominant in the TIR and yield broadband emittance (Fig. S5).

Among materials with lower $\kappa$ in the TIR, many are unsuitable due to poor weatherability (e.g., metal halides), strong UV or visible absorption (e.g., Zinc Sulfide and Germanium), or high cost (e.g., diamond). We therefore focused on lower polyolefins, including polyethene (PE), polypropene (PP), and poly(4-methyl-1-pentene) (PMP), which largely satisfy our previously defined requirements while also being readily available and highly processable (**Figure 3**a and Fig S6). Owing to its simple molecular structure, PE exhibits only a few weak infrared vibrational modes, resulting in $\kappa$ of $10^{-4}$–$2\times10^{-3}$ across much of the TIR band.[52] In PP and PMP, saturated hydrocarbon functional groups give rise mainly to vibrational modes in the LWIR, while $\kappa$ remains as low as $10^{-4}$ over much of the non-LWIR wavelengths.[12] Although such properties of polyolefins have been leveraged in TIR-transmissive designs,[52–55] their use for achieving simultaneous solar-to-TIR ultrabroadband scattering and selective LWIR emission has not been explored. Here we therefore envisioned heterogeneous, disordered media with polyolefins ($n$~1.5) as the high-index component and air ($n$=1) as the low-index component.

### *Ultrabroadband-Scattering PP Underlayer*

We first investigated scatterer geometries that could deliver the desired ultrabroadband solar-to-TIR scattering (Note S2 and Fig. S7). For both PP fibers and particles, $d$ of ~4–16 μm were found to simultaneously support strong TIR scattering and appreciable LWIR absorption, while smaller sizes of $d$~0.5–1 μm enabled efficient solar scattering but negligible TIR scattering (**Figure 3**b and Fig. S8). PE scatterers with comparable sizes exhibit similar scattering behavior but have lower $Q_{abs}$ in the LWIR (Fig. S9). Although this may be advantageous for ultrabroadband reflectance, our aim here is selective LWIR emission, and we therefore focus primarily on TIR-scattering media based on micro-sized PP scatterers.

As a representative case, we then performed FDTD simulations on TIR-scattering media comprising randomly packed PP scatterers with a normal size distribution of 10±6 μm and a volume fraction of ~0.5 (**Figure 3**c, Note S3, and Fig. S10). The normalized electric field distribution profiles and reflection spectra (**Figure 3**d, e) clearly show that, under unpolarized TIR light ($\lambda$~4–20 μm), the medium strongly reflects non-LWIR wavelengths by multiple scattering ($\lambda$~4–6.5 μm and 13–20 μm), while exhibiting lower reflectance and higher absorptance in the LWIR (e.g., $\lambda$~10 μm). Higher $R(\lambda)$, primarily in the far-infrared (FIR), can be achieved by further increasing the medium thickness $t$, although the improvement becomes marginal beyond ~500 μm. This TIR reflectance arises primarily from the large scatterer size, as reducing the size to ~2.0±1.6 μm leads to substantially weaker reflection in the FIR (Fig. S11). Notably, three-dimensional simulations of a microporous PP medium reveal similar optical behaviors (Fig. S12). Collectively, these results demonstrate that the disordered media with tailored scattering architecture and material composition can enable broadband multiple scattering, holding promise to serve as the foundation for realizing either ultrabroadband reflection or selective LWIR emission without metals.

The simulations also support our conjecture that, for a TIR-scattering medium comprising disordered micro-sized scatterers, strong backscattering relative to absorption leads to substantial reflectance at LWIR wavelengths where $\kappa$ remains limited, thereby capping the maximum attainable $\varepsilon_{LWIR}$. This is evident from the observation that increasing $t$ from 300 μm to 1000 μm does not reduce the LWIR reflectance (**Figure 3**d) or increase the absorptance at some wavelengths. Also, LWIR absorptance is notably lower than that of a non-scattering effective medium with identical thickness (**Figure 3**f and Fig. S13). Simulations in visible wavelengths further corroborate the hypothesized limitation of large micro-sized scatters in the geometric regime ($d \gg \lambda \sim 0.5$ μm) (**Figure 3**g), as the spectral reflectance decreases notably towards shorter wavelengths despite the very low intrinsic loss ( $\kappa < 10^{-5}$).

### *Solar-scattering and LWIR-emitting PMP Upper layer*

Guided by the proposed layered architecture to overcome these limitations (**Figure 2**d), we then examined the optical effect of adding a submicron-scattering upper layer onto the PP-based TIR-scattering medium, in both the solar and TIR regimes. For this layer, we chose PMP, which in bulk form is solar-transparent and selectively LWIR-emissive like PP, but has some complementary LWIR absorption peaks (**Figure 3**a and Fig. S6). Their combination could therefore potentially deliver a higher and more spectrally uniform $\varepsilon_{LWIR}$ than the designs based only on PP or PMP.

In the solar wavelengths, we found that placing a 100-μm-thick submicron-scattering layer on a 1000-μm optically thick TIR-scattering medium can increase the average visible reflectance by ~0.16 (**Figure 3**g and Fig. S14). By contrast, mixed-scatterer designs yielded very limited improvement of ~0.04. In the LWIR and FIR, where $d \ll \lambda$ in the upper layer, the negligible scattering and intrinsic emittance of PMP augment and spectrally complement the LWIR emission of the PP underlayer. Moreover, the PMP-air medium acts as an effectively low-index, weakly scattering layer that reduces Fresnel reflection at the air–medium interface. Although these effects also lower the non-LWIR reflectance to an extent, the overall effect is favorable: with a 500-μm-thick and 50%-volume fraction nanoporous PMP top layer, $\varepsilon_{LWIR}$ increases from 0.59 to 0.88 (**Figure 3**h and Fig. S15), and $\eta$ reaches ~1.45 within the simulated 4–20 μm range. By contrast, mixing sub-μm PMP scatterers with ~10 μm-sized PP scatterers yield significantly lower non-LWIR reflectance but higher LWIR reflectance. We note that the performance may be further improved by more precisely tailoring the scattering medium, for example by tuning the scatterer size distribution and volume fraction of the PP TIR-scattering underlayer. (Fig. S16).

## Material Fabrication and Characterization

Having established a theoretical basis for the polyolefin-based layered, multiscale scattering medium, we considered pathways to fabricate it at scale. Promisingly, microfibrous and porous polymers, including polyolefins, are already manufactured at scale for industrial products like separators, filters, and fabrics. Our initial study of such commercially available products showed that few variants indeed exhibit modest, but hitherto unexploited TIR scattering (Fig. S17). In principle, processes like melt extrusion and spinning used to manufacture such variants could be tailored to create microfibrous polyolefins with fiber sizes optimized for scattering TIR radiation.[56] Here in this work to demonstrate a proof of concept with a potential for immediate impact, we custom-ordered a melt-blown microfibrous polypropylene (μPP) fabric and use it for subsequent investigations.[57] While not precisely tailored for optimal performance, the nonwoven μPP fabric has a suitable microstructure, featuring randomly oriented PP fibers with diameters ranging from 2 to 18 μm (**Figure 4**a-b and Fig. S18). Optical measurements show that the μPP, at a thickness of ~500 μm, is indeed an ultrabroadband UV-to-FIR reflector and a wideband LWIR emitter, with $R_{solar}$~0.91, $\varepsilon_{LWIR}$ ~0.51, $\varepsilon_{non\text{-}LWIR}$~0.33 at near-normal incidence (**Figure 4**c).

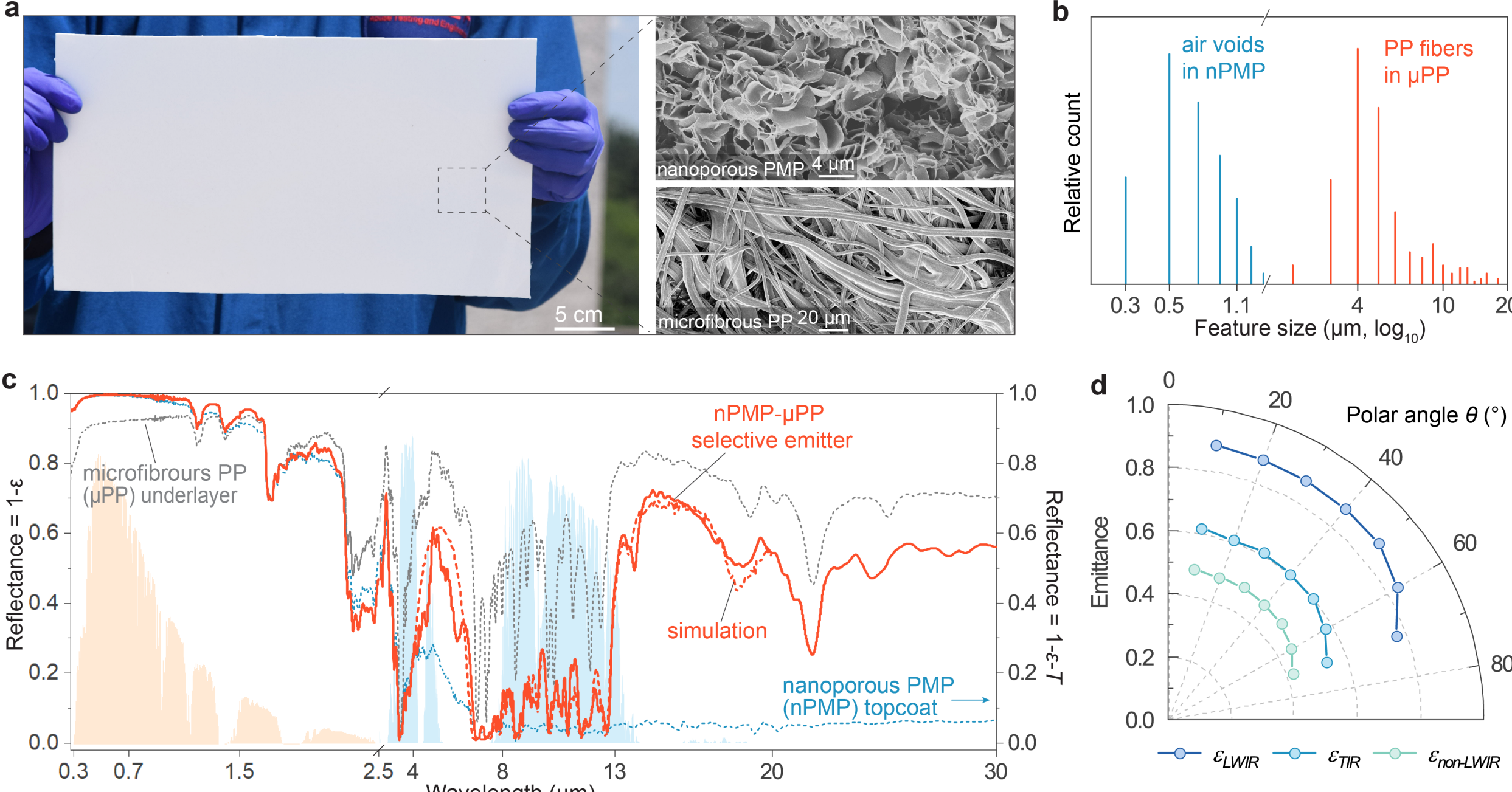

**Figure 4. Characterization of nPMP-μPP.** (**a**) Photograph of a ~40X20 cm² nPMP-μPP selective emitter sample and the micrographs of the nPMP topcoat (showing random PMP lamellae and air voids) and the nonwoven μPP fabric underlayer (showing random PP microfibers). (**b**) Size distributions of air voids within the nPMP and fibers within the μPP. (**c**) Measured spectral reflectance the nPMP-μPP selective emitter at near-normal incidence, and the constituent 500-μm-thick nPMP topcoat and the 500-μm-thick μPP underlayer. The simulated TIR reflectance of a 500-μm-thick nPMP on 500-μm-thick μPP is also plotted as comparison. Plotted against ASTM G173 Global solar spectrum (yellow) and the TIR atmospheric transmittance for mid-latitude summer conditions (light blue). (**d**) LWIR, TIR and non-LWIR emittance of the nPMP-μPP selective emitter as a function of angle. $\varepsilon$: emittance; $T$: transmittance.

However, strong TIR scattering by the PP fibers limits μPP's $\varepsilon_{LWIR}$. Besides, solar reflectance, while high, falls in the shorter solar wavelengths (Fig. S19), as predicted by our theoretical analysis. Thus, guided by the optical design, we experimentally deposited a nanoporous PMP (nPMP) topcoat on the μPP using a scalable phase inversion (immersion-precipitation) method.[58] The process involves casting a film of a PMP-cyclohexane dope solution onto the substrate, followed by immediate immersion in isopropanol as the nonsolvent (Methods). The rapid solvent-nonsolvent exchange microstructures the PMP into petal-like lamellae, creating within minutes numerous interstitial voids with an average size of ~0.5 μm (**Figure 4**a-b and Fig. S20). These features act as Mie scatterers of sunlight, causing a 160-μm-thick nPMP film to attain $R_{solar}$ >0.91 (Fig. S21), and >0.95 when placed on the μPP (Fig. S22). Thermally bonding the μPP and nPMP enhances mechanical properties of the layered architecture (Fig. S23). By comparing the optical characteristics at different nPMP thicknesses (Fig. S22), the final nPMP-μPP selective emitter was chosen to comprise a ∼500 μm-thick nPMP topcoat on a ∼500-μm-thick μPP underlayer, which yields high $R_{solar}$=0.97±0.01, $\varepsilon_{LWIR}$=0.88±0.02, and notable $\eta$=1.42±0.04 at near-normal incidence (**Figure 4**c and Fig. S22). The well agreement of the measured TIR reflection spectrum with our optical simulation result verifies that the size scales and materials of scatterers can be synergistically optimized to realize complex spectral functionalities. Moreover, because of the scattering-based optical mechanism and low effective index of the nPMP topcoat, the selective emittance extends across a wide angular range (**Figure 4**d), yielding hemispherical $\varepsilon_{LWIR}$ of ∼0.85 and $\eta$ of ∼1.45.

## Evaluation of Thermal Performance

The high solar reflectance and selective LWIR emittance of nPMP-μPP enable it to achieve notable radiative thermoregulation relative to broadband emitters across a range of scenarios. We substantiated this through a series of field tests comparing the steady-state temperatures of and heat flows through nPMP-μPP samples relative to two broadband controls (Fig. S24). The first control is a commercial white paint, served as a typical benchmark ($R_{solar}$=0.83, $\varepsilon_{LWIR}$=0.94, $\eta$=1.03). The second is a broadband RC with the same underlayer as nPMP-μPP, but with nPMP replaced by an equivalent thickness of expanded polytetrafluoroethylene (ePTFE), yielding nearly identical $R_{solar}$=0.96 and $\varepsilon_{LWIR}$=0.88 to nPMP-μPP, but a broadband emittance ($\eta$=0.99). Comparison with the ePTFE-μPP therefore parses out the effect of selective emittance. The tests were conducted with samples in sky-facing or vertical orientations and in urban or semi-urban environments, during summer days with and without direct sunlight, and during winter nights. Notably, rather than being shielded from thermal convection by windshields as in many previous studies, all samples were fully exposed to ambient conditions. Furthermore, each sample was mounted on a steel backplate to increase thermal capacity and insulated from the rear to suppress parasitic heat flows (**Figure 5**a).

### *Skyward Radiative Cooling Under Unconducive Sky Conditions*

Sky-facing radiative cooling tests were conducted in Princeton (New Jersey, USA) during the summer of 2025, when strong sunlight, sporadic cloud cover, warm ambience, periodic winds, and high total precipitable water (TPW) made sub-ambient radiative cooling particularly challenging. Over a 4-hour period on June 5 afternoon with a peak/mean solar irradiance $I_{solar}$ of 1000/840 W m$^{-2}$, TPW of ∼28 mm and average wind speed of ∼4 m s$^{-1}$, the nPMP-μPP selective emitter stayed ∼4.9/4.0°C (peak/mean) cooler than the white paint, 2.4/1.6°C cooler than ambient air, and 0.8/0.5°C cooler than the ePTFE-μPP broadband RC (**Figure 5**b,c). Notably, this cooling was achieved when the combination of high TPW and scattered clouds partially closed the atmospheric transparency window and raised the radiant temperature $T_{radiant}$ of sky to within 12°C of the ambient temperature $T_{amb}$. As a benchmark, under such conditions the maximum attainable average radiative cooling power $\bar{P}_{cooling}$ for a near-ideal broadband RC ($R_{solar}$=0.97, $\varepsilon_{LWIR}$=1, $\eta$=1) is ∼47 W m$^{-2}$ when it is at $T_{amb}$. We calculated $\bar{P}_{cooling}$ of the samples tested (Note S4): whereas the white paint achieved a net radiative heat gain of ∼57 W m$^{-2}$, the broadband RC and nPMP-μPP samples achieved average radiative cooling of ∼27 and ∼40 W m$^{-2}$ respectively, under steady-state temperatures (**Figure 5**c). If the emitters were held at $T_{amb}$, the $\bar{P}_{cooling}$ of nPMP-μPP would have been ~50 W m[-2], 17 W m[-2] higher than that of ePTFE-μPP.

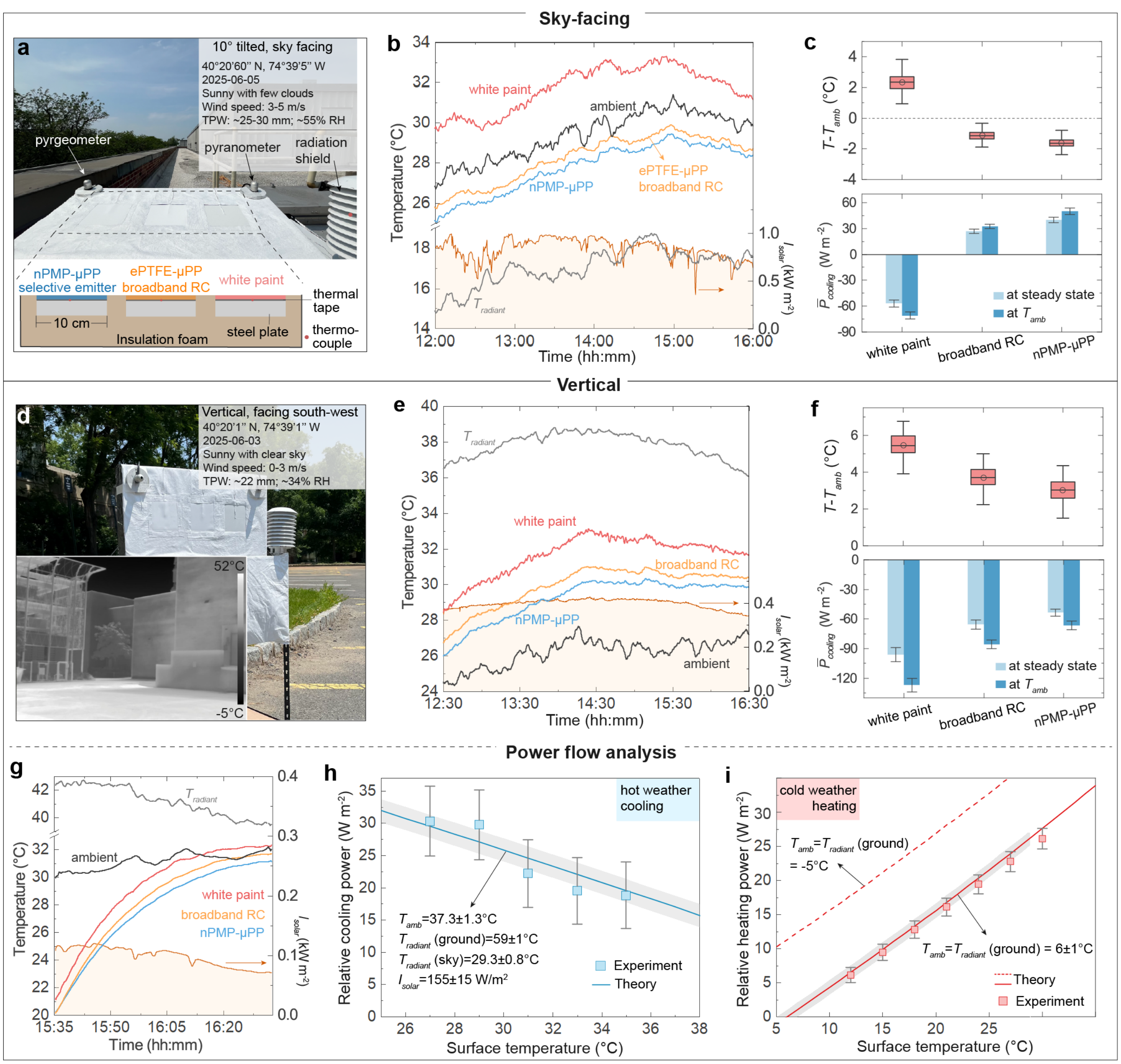

**Figure 5. Field tests of thermoregulation performance.** (**a-c**) When samples are horizontally placed facing the sky: (**a**) photograph and schematic of the field test setup; (**b**) real-time temperature data, TIR irradiance expressed in radiant temperature $T_{radiant}$, and solar irradiance $I_{solar}$; (**c**) box-plot of the sample temperature $T$ relative to ambient temperature $T_{amb}$, and the average radiative cooling power $P_{cooling}$ when samples are either steady-state or assumed to be at $T_{amb}$. (**d-f**) When samples are vertically placed facing both sky and terrestrial environment: (**d**) photograph of the field test setup and a thermal photo showing the radiant environment in the field of view; (**e**) real-time temperature data, TIR irradiance expressed in $T_{radiant}$, and solar irradiance $I_{solar}$; (**f**) box-plot of the sample temperature $T$ relative to ambient temperature $T_{amb}$, and the average radiative cooling power $P_{cooling}$ when samples are either steady-state or assumed to be at $T_{amb}$. (**g-i**) Differential power flow analysis in vertical orientation: (**g**) field test results when samples were vertically placed in a hot terrestrial environment without direct sunlight and were heated by the environment; (**h**) radiative heat gains prevented (i.e., relative cooling) in a hot environment; (**i**) radiative heat losses prevented (i.e., relative heating) in a cold environment by the nPMP-μPP compared to the ePTFE-μPP. Note: $T_{amb}$ is averaged with a moving 5-minute window, while the other temperature data are one-minute average.

Additional tests conducted under other weather conditions yielded similar results, including on 23 June during a heat-dome event[59] that produced very hot ($T_{amb}$ up to ~37°C) and humid weather (TPW up to ~35 mm) (Fig. S25). Between 08:00 and 16:00 on that day, while the maximum attainable $\bar{P}_{cooling}$ of a near-ideal broadband RC was only ~25 W m$^{-2}$, the nPMP-μPP selective emitter attained average cooling of 4.3°C, 2.1°C and 0.4°C relative to white paint, ambient air, and broadband RC, respectively. Overall, these results highlight how the high solar reflectance and selective LWIR emittance of nPMP-μPP effectively suppresses radiative heat gains, enabling sky-facing variants to remain sub-ambient even when environmental conditions are non-ideal for radiative cooling.

### *Radiative Thermoregulation in Vertical Orientation*

While radiative cooling of sky-facing emitters has been long explored, the ability of selective emitters to thermoregulate vertical surfaces relative to broadband emitters – i.e., relative cool in hot weather and heat in cold weather – has only been recently elucidated.[12,9,13] When vertically oriented and facing the hot thermal environments experienced by building facades and human bodies in urban settings, the cooling benefits of the nPMP-μPP selective emitter become even more pronounced. In contrast to the sky-facing configuration, where the $T_{radiant}$ was lower than $T_{amb}$ (**Figure 5**a), vertically oriented samples were exposed to warm ground and surrounding buildings, leading to elevated TIR irradiance with $T_{radiant}$ ~12°C higher than $T_{amb}$ during the field test (**Figure 5**d and Fig. S26). This corresponds to over 70 W m$^{-2}$ of net TIR heat load, forcing the highly solar-reflective but broadband emissive ePTFE-μPP to remain ~3.7°C above $T_{amb}$ on average (**Figure 5**e). The white paint, owing to its lower solar reflectance, was ~5.5°C above $T_{amb}$. By reducing TIR terrestrial heat gain from the environment relative to the broadband RC, and also solar heat gain relative to the white paint, nPMP-μPP stayed ~0.7°C and ~2.5°C cooler relative to the two emitters (**Figure 5**f). Similar trends were also observed under other warm outdoor conditions (Fig. S27). The relative cooling performance in such conditions is similar to those of metallized selective emitters.[12,13]

Because vertical surfaces receive substantially more non-LWIR irradiation but less direct solar irradiation than horizontal surfaces, the selective LWIR emission of nPMP-μPP plays a more critical role in thermoregulation. This was demonstrated in a test with samples exposed to a hot terrestrial environment but only diffuse sunlight ($I_{solar}$~100 W m$^{-2}$), showing that only nPMP-μPP remained cooler than ambient air after 1 h of heating by the surrounding environment (**Figure 5**g). Analysis of the temperature rise confirmed that nPMP-μPP absorbed notably less radiant heat (Fig. S28), and since solar absorption differences were less than 2 W m$^{-2}$, the observed cooling is primarily attributed to reduced absorption of non-LWIR terrestrial irradiation.

We further performed field tests and detailed power flow analysis to quantify this effect (Note S5). Under a hot condition, with $T_{amb}$ of ~37.3°C and the $T_{radiant}$ of ground reaching ~60°C (Fig. 29), both theoretical calculations and experimental measurements showed that nPMP-μPP attained a relative cooling power of ~30 W m$^{-2}$ at 27°C and ~ 17 W m$^{-2}$ at $T_{amb}$ compared with the ePTFE-μPP broadband RC (**Figure 5**h). Notably, at representative human-body surface temperatures of 32-35 °C, the relative cooling still exceeded 20 W m$^{-2}$. Moreover, under cold conditions such as winter nights, when building surfaces and human bodies lose substantial heat to the environment,[60] nPMP-μPP's low emittance in the non-LWIR also enabled reduced radiative heat loss and relative heating effect (**Figure 5**i and Fig. S30). For building surfaces, which can be as much as 10-15°C warmer than the terrestrial environment in the winter, we observed relative heating of 10-15 W m$^{-2}$ when $T_{amb}$ was ~6°C, and at human-body surface temperatures, the relative heating was ~30 W $m^{-2}$. Notably, these results are largely consistent with theoretical predictions. The magnitude of the relative cooling or heating power increases with the temperature difference between the thermoregulating surface and the surrounding radiant environment, by ~1.2 W $m^{-2}$ per degree Celsius.

## Potential Energy Impact

### *Deployability*

Our work demonstrates that a selective LWIR emitter, which provides superior passive thermoregulation compared to more common broadband emitters, can be realized entirely using scattering media. This may offer broad utility and some advantages over conventional designs, including low cost and high scalability, thereby facilitating the practical deployment of spectrally selective radiative thermal management. Preliminary analysis of the proof-of-concept nPMP-μPP suggests that the cost can be constrained to ~US$3 $m^{-2}$ (Table S3 and S4). This is notably lower than the cost of many other selective LWIR emitters,[4,13,25,30] and could be further reduced when totally using established industrial manufacturing process for microfibrous polymers that are solvent-free (Fig. S31),[57] being comparable with other scattering-media designs like paints and fabrics.[28,54,61] Moreover, the appearance of scattering media can be readily tuned by incorporating colorants while maintaining LWIR selectivity (Fig. S32), while eliminating metals can reduce glare in outdoor applications and may mitigate weathering-induced degradation of optical properties. The porous morphology and metal-free composition also impart flexibility and breathability without compromising waterproofness, expanding the range of potential applications (Fig. S33). Although the focus of this work is on optical design, we note that a key engineering challenge is to improve material durability under prolonged outdoor exposure in some relevant applications, as polyolefins are often susceptible to chemical damage by UV irradiation. An accelerated UV aging of the nPMP-μPP under a very high intensity of 350 W $m^{-2}$ for one week shows little impact on its optical performance (Fig. S34). Potential improvement pathways include incorporating UV stabilizers that are also spectrally selective or have little impact on TIR optical properties.[62,63]

Along with the demonstrated radiative cooling and thermoregulation performance relative to traditional designs, these attributes present possibilities for all-season building energy savings and improving thermal comfort, as discussed below.

### *Building Envelopes for Energy Efficiency*

Cooling-energy savings associated with highly solar-reflective, LWIR-emissive roofs are well documented.[26,64,65] While a selective emitter does enable better radiative cooling than a broadband emitter, the improvement is often modest at best for roofs.[33] Nonetheless, the overall cooling benefit remains substantial, and in the case of our design, achievable by overcoming the design limitations and cost of metal-based selective LWIR emitters reported thus far. A recently demonstrated, and potentially more impactful opportunity lies in the thermoregulation of facades that see the terrestrial environment,[12] where the nPMP-μPP's optical properties provide both cooling in hot weather and heating in cold weather relative to broadband emitters (**Figure 5**h, i). Also, walls are often less insulated and present larger surfaces than roofs. Therefore, here we focus on the energy-saving potential for facade thermoregulation using a building thermal model with the capability of simulating spectral selective facades (Note S6).[12,60] We evaluated the energy savings for a mid-sized building in three cities representing distinct climate regimes (**Figure 6**a). The building was assumed to have either wooden walls with R13 insulation, as is common in the US, or brick walls, as is often the case in the global south.

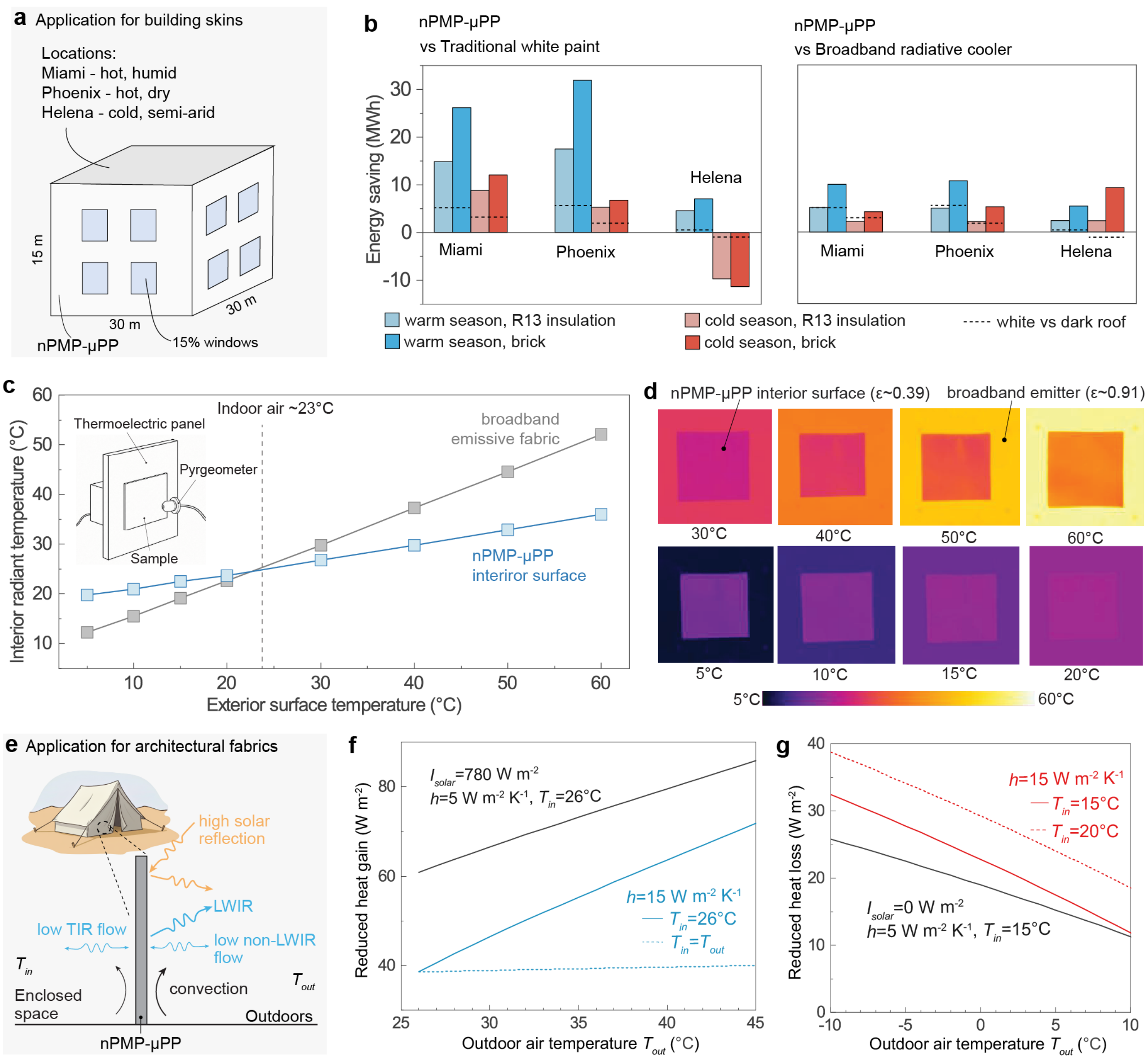

**Figure 6. Potential impact on improving energy saving and thermal comfort.** (**a**) Schematic of the mid-size residential building for modelling energy saving performance on walls. (**b**) Energy savings achieved in three cities representing distinct climate regimes, relative to a typical white paint and a broadband radiative cooler. (**c**) Effective radiant temperature from the indoor-facing surface of nPMP-μPP and a broadband emissive fabric under controlled exterior surface temperatures. The sample was attached to a thermoelectric panel whose temperature was set to selected values from 5 to 60°C, representing the exterior surface temperature of the samples in different outdoor conditions. A pyrgeometer that captures TIR radiation was positioned facing the sample to measure its thermal radiosity. (**d**) Infrared thermal images of the experimental setup at different thermoelectric panel temperatures, showing that nPMP-μPP maintains a relatively stable radiant temperature on its interior-facing surface despite large temperature variations in exterior surface. (**e**) Schematic of heat flows through the nPMP-μPP as an architectural fabric that enables strong reflection of solar heat, reduced non-LWIR heat exchange with terrestrial environment, and reduced TIR heat exchange with the indoor environment.

This significantly reduces (**f**) heat gains by the enclosed space in hot weather (e.g., summer daytime), and (**g**) heat loss in cold weather (e.g., winter nighttime). $h$: outdoor convective coefficient; $T_{in}$: indoor temperature.

When benchmarked against typical white surfaces ($R_{solar}$=0.86, $\varepsilon_{LWIR}$=0.9, $\eta$=1), exterior wall envelopes made from nPMP-μPP reduce cooling and heating loads across warm and hot climates, enabling ~22-39 MWh annual energy savings and $CO_2$ reductions exceeding ~5-9 tons (**Figure 6**b). This often surpasses the well-documented benefits associated with painting dark roofs white.[64] In cold-climates regions like Helena, the ultrahigh $R_{solar}$ of nPMP-μPP introduces a penalty due to overcooling in winter; however, it still outperforms broadband RCs with comparable $R_{solar}$ ($R_{solar}$=0.97, $\varepsilon_{LWIR}$=0.9, $\eta$=1) (**Figure 6**b), highlighting the intrinsic advantage of spectrally selective TIR emission. Crucially, with the design of layered scattering media, $R_{solar}$ and $\varepsilon_{TIR}$ can be decoupled and tuned, e.g., by introducing colorants to reduce $R_{solar}$ while largely preserving TIR properties (Fig. S32), and by adjusting the dosage of the top layer to adjust $\varepsilon_{LWIR}$ (Fig. S22). Simulations show that a darker, LWIR-selective variant ($R_{solar}$=0.86, $\varepsilon_{LWIR}$=0.88, $\eta$=1.42) retains notable energy savings across both hot and cold climates (up to ~10 MWh for brick walls, Fig. S35). Moreover, since the μPP underlayer has a selective but lower $\varepsilon_{LWIR}$ of 0.51 and $\varepsilon_{TIR}$ of 0.39, if used solely by itself to enhance the reduction in TIR heat loss in heating-dominated conditions, it may substantially improve energy efficiency by ~10 MWh with a modest R13 insulation (Fig. S35). These results indicate our design could extend scattering media to be a more flexible platform for engineering materials with application-specific radiative regulation functionalities.

### *Wearable and Architectural Fabrics*

Radiative thermoregulation can also directly enhance human thermal comfort in minimally sheltered settings or under outdoor exposure.[66] The optical design proposed in this work is compatible with polymeric fabrics,[67] thus offering a promising paradigm for thermoregulation textile engineering. While previous studies have established that textiles with high $R_{solar}$ can effectively reduce heat stress under sunlight,[24,43] their susceptibility to broadband TIR heat in urban settings has only been recently noticed.[12,13] Our research demonstrates that nPMP-μPP, beyond solar reflection, can further reduce undesired radiant fluxes at temperatures characteristic of clothed human bodies in both hot and cold urban environments, by ~20 and ~30 W $m^{-2}$, respectively (**Figure 5**h, i). These are about 20% of human body metabolic rate.[68]

The potential of nPMP-μPP goes also beyond being an exterior layer of building envelopes. Notably, because the interior-facing side of nPMP-μPP exhibits a much lower emittance than that of typical nonmetallic materials, as a standalone envelope separating indoor and outdoor environments, it can maintain a relatively stable interior radiant temperature despite large variations in exterior surface temperature (**Figure 6**c, d). This can substantially suppress radiative heat flows to/from the interiors compared to broadband-emissive fabrics. For instance, for an indoor environment at 23°C, the nPMP-μPP may exhibit an effective radiant temperature ~10°C lower than typical fabrics when their exterior surfaces are ~40-50°C, representing hot environments (**Figure 6**c). When they are much colder, the nPMP-μPP likewise shows a warmer radiant temperature (**Figure 6**c, d). This highlights its potential for applications where direct contact with the thermoregulated object is not required and where outdoor temperatures are either much higher or much lower than the desired thermally comfortable indoor environment.

Typical examples are architectural fabrics, which are used to provide basic shelter with minimal material usage (e.g., in tents and emergency shelters), yet conventional designs offer poor thermal insulation due to strictly limited thickness (~1 mm) and thermal mass.[69] We performed heat-flow calculations to gauge the thermal performance of nPMP-μPP in such scenarios (**Figure 6**e and Note S7). Under hot conditions with solar irradiance of ~780 W m$^{-2}$ and ground temperatures 20°C above outdoor $T_{amb}$, a vertically oriented nPMP-μPP reduces heat gain into temperature-controlled indoors by ~40-80 W m$^{-2}$ (**Figure 6**f), indicating substantial cool energy saving potential. Even without indoor temperature control and indoor temperature $T_{in}$ reaches $T_{amb}$, it still reduces heat gain by ~39 W m$^{-2}$, a >80% reduction over that of conventional fabrics, and comparable to increasing the thickness of conventional fabrics by ~75 mm. During cold nights, when $T_{amb}$ drops to 0 °C and $T_{in}$ is maintained at the minimum recommended 15°C, nPMP-μPP could also reduce heat loss by >23 W m$^{-2}$ (**Figure 6**g).

## Conclusions

We developed a layered, multiscale scattering medium that delivers ultrabroadband reflection from the ultraviolet to the far-infrared while enabling band-selective absorption, without relying on metals or other substrates. Leveraging this architecture, we demonstrate a fabric-like selective LWIR emitter that can radiatively insulate underlying surfaces from non-LWIR radiative heat gain (and loss) while maintaining efficient LWIR exchange with the sky. Optical and thermal analysis and experiments validate the designed spectral response and show improved passive radiative thermoregulation across diverse scenarios relative to broadband emitters. By the nature of the low cost, established scalable manufacturing, and tunability of scattering media, this strategy holds promise for advancing passive radiative heat control for building energy savings and human thermal comfort. More broadly, our results show optical selectivity over an ultrabroad waveband without the aid of bulk reflectors, and mark an advancement over prevailing mixture-based paradigm in scattering-media design[23,26,49] by highlighting the advantage of a multiscale, layered organization of scatterers based on their intrinsic optical properties. The same design principle could be applicable to a broad range of systems that harness multiple elastic waves scattering in complex media for controlling broadband reflection and absorption, such as coatings, energy harvesting, and camouflage.

## Methods

### *Sample fabrication*

The nanoporous PMP (nPMP) film was made by an immersion-precipitation phase inversion process. First, solid PMP were dissolved in cyclohexane in a 1:11 mass ratio at ~70°C under continuous stirring until getting a clear and homogenous dope solution. Subsequently, the dope solution was blade-coated on a clean glass substrate, which was then immersed in an isopropanol bath at ~40°C. After ~10 minutes when the clear film became turbid, it was taken out and dried at room temperature to become a super white nanoporous PMP film. The microfibrous PP nonwoven textiles (μPP) were made by melt-blowing, and custom ordered from Changzhou Jinchun Environmental Protection Technology Co., Ltd. The nPMP-μPP was then made by thermally bonding the nPMP film on top of the μPP (Fig. S23). The optimal thickness of the nPMP film was selected to be ~500 μm and was used for field tests and further analysis.

The ePTFE-μPP broadband radiative cooler was made by thermally laminating a 500-μm-thick expanded polytetrafluoroethylene (ePTFE) on top of the μPP textile with a thin PE adhesion layer in the middle. This served as a broadband radiative cooler control with a nearly identical solar reflectance. A commercial white paint (BEHR Premium White) was spray-coated on steel with over 300 μm thickness, used as a typical benchmark.

### *Optical simulation*

Finite-difference-time-domain (FDTD) electromagnetic simulations were conducted to study the optical behaviors of disordered scattering media, which were represented by random, non-overlapping packings of polydisperse disks (for 2D simulations) or spheres (for 3D simulations). A plane wave was used as the incident light source and was normally incident on the structure. Two power monitors were placed above the light source and below the structure, respectively, to record the reflectance and transmittance. Perfectly matched layer (PML) boundary conditions were applied to the boundaries perpendicular to the incident light direction, while periodic boundary conditions were imposed in the lateral directions along the structural boundaries. Both S- and P-polarized light sources were separately used, and their averaged results were used to represent unpolarized illumination.

For comparing the effects of scatterers with different size ranges and $\kappa$ values on the optical properties of scattering media, we performed 2D simulations. The 2D ensembles were generated by random packing of circular disks within prescribed size ranges, forming structures with a thickness of 1100 μm and a width of 200 μm. At this thickness, the scattering media were nearly opaque for light in most solar and TIR wavelengths. For the simulations of PP-based scattering media, in addition to assigning the measured refractive index data of PP to the scatterers, we randomly generated 10 structures for each case to further approximate real structures with a nearly infinite lateral dimension. These structures had nearly identical scatterer sizes, volume fractions, and total thicknesses. The averaged results from the 10 structures were used as the final results. More details can be found in Notes S2 and S3.

### *Optical and morphological characterizations*

Spectral properties of all samples in this work were determined separately in the wavelength range of 0.28-0.9 μm, 0.9-1.85 μm, and 1.85-23 μm. For the first range, measurement was taken using a UV-VIS Cary 5000 Spectrometer equipped with an integrating sphere. For the second range, measurement was taken using a Fourier Transform Infrared (FTIR) spectrometer (Bruker Invenio) equipped with a 100-mm integrating sphere (4P4, Thorlabs) and a Germanium photodiode (SM05PD6A, Thorlabs). For the third range, the FTIR spectrometer was equipped with a gold integrating sphere along with a mercury cadmium telluride (MCT) detector. For the 0.28-1.85 μm wavelengths, a diffuse reflectance standard (Spectralon SRS-99-010, Labsphere) was used as a reference, while for larger wavelengths, a diffuse gold coating was used. Moreover, in the far infrared wavelengths ($\lambda \sim$20-50 μm), the specular reflectance was measured by the FTIR spectrometer with DTGS detectors and using a silver mirror as reference. The results were scaled and patched with the integrating sphere measurements in the overlapped wavelengths to estimate the total reflectance in far infrared wavelengths.

With the measured spectral total reflectance $R(\lambda)$, the average solar reflectance $R_{solar}$ is calculated by $R_{solar} = \frac{\int_{\lambda_1}^{\lambda_2} I_{AM1.5}(\lambda)\cdot R(\lambda)d\lambda}{\int_{\lambda_1}^{\lambda_2} I_{AM1.5}(\lambda)d\lambda}$, where $I_{AM1.5}(\lambda)$ is the ASTM G173 global solar intensity spectrum at the air mass of 1.5, and the wavelength range is 0.28-2.5 μm. The average thermal emittance is calculated by $\varepsilon = \frac{\int_{\lambda_1}^{\lambda_2} I_{BB}(\lambda)\cdot(1-R(\lambda))d\lambda}{\int_{\lambda_1}^{\lambda_2} I_{BB}(\lambda)d\lambda}$, where $I_{BB}(\lambda)$ is the spectral radiance emitted by a blackbody at 30°C. For average LWIR emittance $\varepsilon_{LWIR}$, the wavelength range is 8-13 μm; for average TIR emittance $\varepsilon_{TIR}$, the wavelength range is 2.5-50 μm unless otherwise stated; with them, LWIR selectivity is determined by $\eta = \frac{\varepsilon_{LWIR}}{\varepsilon_{TIR}}$. We further note that the emittance values correspond to the case where the sample is placed onto a perfectly absorptive substrate, with all reflectance arising from the sample itself.

Thermographs were taken using a FLIR T865 thermal camera working at a wavelength range of 7-14 μm. Photographs were taken using a Nikon D3300 DSLR camera. The microstructure of the nPMP films and μPP fabrics was imaged using a scanning electron microscope (Verios 460, FEI).

### *Field tests*

All samples (~10 cm × 10 cm) were mounted on steel blocks with thermally conductive tapes. The blocks were then fitted into recesses in an expanded polystyrene insulation panel (Foamular NGX by Owens Corning, ~51 mm thick, thermal conductivity ~0.029 W $m^{-1}$ $K^{-1}$), with all exposed insulation surfaces covered with white Tyvek over aluminum foils to suppress solar heating and terrestrial radiative heating (Figure 4a). The front surfaces of the samples were flush with the panel and fully exposed to air. Sample temperatures were measured using K-type thermocouples embedded between the sample and the steel block, and the ambient air temperature was monitored by a thermocouple placed inside a solar radiation shield. Solar irradiance was measured using a pyranometer and TIR irradiance from the environment was measured by a pyrgeometer. They were also both mounted on the insulation panel, away from the view of the samples but at the same tilt angle and orientation as the samples. During the experiments, the panel was positioned on a 1.2 m-high stand and tilted at either 10° towards the south (sky-facing test, to maximize solar irradiance on the samples) or 90° (vertical test). The surface of the stand was painted white to minimize heating of the surrounding air near the samples.

Furthermore, the methods used for building energy saving simulations and heat transfer analysis of architectural fabrics are explained in Notes S6 and S7.

## Acknowledgements

The study was funded by start-up funds and Sustainability of our Planet funds provided by the School of Engineering and Applied Science at Princeton University. Microscopy was performed in the Imaging and Analysis Center at Princeton University.

**Author Contributions**: J. M. and Z.L. conceived the concept, performed optical simulations and experiments, and wrote the manuscript. Z.L. fabricated the materials, performed the tests and data analysis. M.D. performed building energy simulations. N.J.V. assisted with materials characterization. J.M. supervised the research project.

**Competing interests:** A provisional US patent application (no. 63/990,394), where J. M. and Z. L. are inventors, has been filed related to this work.

**Data and materials availability:** All data are available in the manuscript or the supplementary materials. Information requests should be directed to the corresponding authors.

**Supplementary information:** Supplementary Notes S1-7, figures S1-35, tables 1-4 and references.